\documentstyle[epsfig]{mn}

\newif\ifAMStwofonts
\ifoldfss
  \ifCUPmtlplainloaded \else
    \NewTextAlphabet{textbfit} {cmbxti10} {}
    \NewTextAlphabet{textbfss} {cmssbx10} {}
    \NewMathAlphabet{mathbfit} {cmbxti10} {} % for math mode
    \NewMathAlphabet{mathbfss} {cmssbx10} {} %  "   "    "
  \fi
  \ifAMStwofonts
    \ifCUPmtlplainloaded \else
      \NewSymbolFont{upmath} {eurm10}
      \NewSymbolFont{AMSa} {msam10}
      \NewMathSymbol{\upi}     {0}{upmath}{19}
      \NewMathSymbol{\umu}     {0}{upmath}{16}
      \NewMathSymbol{\upartial}{0}{upmath}{40}
      \NewMathSymbol{\leqslant}{3}{AMSa}{36}
      \NewMathSymbol{\geqslant}{3}{AMSa}{3E}

      \let\geq=\geqslant 
    \fi
  \fi
\fi % End of OFSS

\ifnfssone
  \newmathalphabet{\mathit}
  \addtoversion{normal}{\mathit}{cmr}{m}{it}
  \addtoversion{bold}{\mathit}{cmr}{bx}{it}
  \newmathalphabet{\mathbfit} % math mode version of \textbfit{..}
  \addtoversion{normal}{\mathbfit}{cmr}{bx}{it}
  \addtoversion{bold}{\mathbfit}{cmr}{bx}{it}
  \newmathalphabet{\mathbfss} % math mode version of \textbfss{..}
  \addtoversion{normal}{\mathbfss}{cmss}{bx}{n}
  \addtoversion{bold}{\mathbfss}{cmss}{bx}{n}
  \ifAMStwofonts
    \ifCUPmtlplainloaded \else
      \UseAMStwoboldmath
      \makeatletter
      \new@mathgroup\upmath@group
      \define@mathgroup\mv@normal\upmath@group{eur}{m}{n}
      \define@mathgroup\mv@bold\upmath@group{eur}{b}{n}
      \edef\UPM{\hexnumber\upmath@group}
      \new@mathgroup\amsa@group
      \define@mathgroup\mv@normal\amsa@group{msa}{m}{n}
      \define@mathgroup\mv@bold\amsa@group{msa}{m}{n}
      \edef\AMSa{\hexnumber\amsa@group}
      \makeatother
      \mathchardef\upi="0\UPM19
      \mathchardef\umu="0\UPM16
      \mathchardef\upartial="0\UPM40
      \mathchardef\leqslant="3\AMSa36
      \mathchardef\geqslant="3\AMSa3E

      \let\geq=\geqslant 
    \fi
  \fi
\fi % End of NFSS release 1

\ifnfsstwo
  \DeclareMathAlphabet{\mathbfit}{OT1}{cmr}{bx}{it}
  \SetMathAlphabet\mathbfit{bold}{OT1}{cmr}{bx}{it}
  \DeclareMathAlphabet{\mathbfss}{OT1}{cmss}{bx}{n}
  \SetMathAlphabet\mathbfss{bold}{OT1}{cmss}{bx}{n}
  \ifAMStwofonts
    \ifCUPmtlplainloaded \else
      \DeclareSymbolFont{UPM}{U}{eur}{m}{n}
      \SetSymbolFont{UPM}{bold}{U}{eur}{b}{n}
      \DeclareSymbolFont{AMSa}{U}{msa}{m}{n}
      \DeclareMathSymbol{\upi}{0}{UPM}{"19}
      \DeclareMathSymbol{\umu}{0}{UPM}{"16}
      \DeclareMathSymbol{\upartial}{0}{UPM}{"40}
      \DeclareMathSymbol{\leqslant}{3}{AMSa}{"36}
      \DeclareMathSymbol{\geqslant}{3}{AMSa}{"3E}

      \let\geq=\geqslant 
    \fi
  \fi
\fi % End of NFSS release 2

\ifCUPmtlplainloaded \else
  \ifAMStwofonts \else % If no AMS fonts
    \def\upi{\pi}
    \def\umu{\mu}
    \def\upartial{\partial}
  \fi
\fi

\title{Models for the Magnitude-Distribution of Brightest Cluster Galaxies}

\author[J. P. Bernstein and Suketu P. Bhavsar]
	{J. P. Bernstein$^1$ and Suketu P. Bhavsar$^{1,2}$ \\
	$^1$  Department of Physics and Astronomy, University of Kentucky, 
	Lexington, KY 40506-0055, USA \\
	$^2$  IUCAA, Pune 411 007, India (on sabbatical leave from 
	U. Kentucky 1999 July - 2000 July)}
\date{Accepted 2000 October 18.
      Received 2000 Septebmer 14;
      in original form 2000 March 23}

\pagerange{\pageref{firstpage}--\pageref{lastpage}}
\pubyear{2000}

\begin{document}

\maketitle

\label{firstpage}

\begin{abstract}
The brightest, or first-ranked, galaxies (BCGs) in rich clusters show a
very small dispersion in luminosity, making them excellent standard 
candles. This small dispersion raises questions about the nature of 
BCGs. Are they simply the extremes of normal galaxies formed via a 
stochastic process, or do they belong to a special class of atypical 
objects? If they do, are all BCGs special, or do normal galaxies 
compete for the first rank? To answer these questions, we undertake a 
statistical study of BCG magnitudes using results from extreme value 
theory. Two-population models do better than do one-population models.
A simple model where a random boost in the magnitude of a fraction of 
bright normal galaxies forms a class of atypical galaxies best describes 
the observed distribution of BCG magnitudes.
\end{abstract}


\section{Introduction}
Among the most luminous bodies in the universe are the brightest, or 
first-ranked, galaxies in rich clusters. These galaxies have absolute 
magnitudes between -21.5 and -23.3 and are among the farthest 
observable objects. In addition, the magnitudes of these brightest 
cluster galaxies (BCGs) are highly uniform, with a dispersion of 0.32 
magnitudes (Hoessel \& Schneider 1985). Their uniformity and large 
luminosity make BCGs excellent standard candles. The uniformity of 
BCG magnitudes raises a particularly important question regarding 
their nature (Peebles 1968; Sandage 1972). Are BCGs simply the 
brightest of a statistical set of galaxies or do they belong to a special 
class of objects? If a special class of galaxies exists, do all clusters 
have special galaxies and are they always first-ranked (Bhavsar 1989)? 
We investigate these questions using extreme value theory (Fisher \& 
Tippett 1928).

\section{Extreme Value Theory}
The motivation for studying extreme phenomena is practical. Many of 
the memorable experiences in our lives can be classified as statistical 
extremes. Examples of maximum extremes are floods, the hottest 
summer temperatures and the lengths of the longest caterpillars. 
Examples of minimum extremes are draughts, stock market crashes 
and the wing-spans of the smallest hummingbirds. Some extremes do 
not effect our lives and others turn them upside down. The desire to 
understand these types of phenomena prompts the study of extreme 
value theory.

Fisher \& Tippett (1928) show that the distribution of statistically 
largest or smallest extremes tends asymptotically to a well-determined 
and analytic form for a general class of parent distributions. Extremes 
drawn from sufficiently large and steeply falling parent distributions 
have this form. One may find the original argument in Fisher \& Tippett 
(1928). Their derivation is reconstructed in greater detail by Bhavsar \& 
Barrow (1985), who apply extreme value theory in an analysis of BCG 
magnitudes. Fisher \& Tippett's result states that the cumulative 
distribution of maximum extremes is given by:
\begin{equation}
        F(x)=e^{-e^{-a(x-x_0)}}.
\label{Eq:eq1}
\end{equation}
This distribution is known as the $Gumbel$ distribution. (For smallest 
extremes, one substitutes $x \to -x$.) From $F$ we may calculate the
differential distribution (or probability density): 
\begin{equation}
        f(x)=ae^{-a(x-x_0)-e^{-a(x-x_0)}},
\label{Eq:eq2}
\end{equation}
where $f(x)$ = $F'(x)$; $x_0$ is the mode of the extremes and 
$a > 0$ is a measure of the steepness of fall of the parent distribution. 
The probability density is normalized to unity. The mean, median 
and standard deviation of the distribution given in Bhavsar \& Barrow 
(1985) correspond to:
\begin{equation}
        <x>=x_0+\frac{0.577}{a}; \; med(x)=x_0+\frac{0.367}{a}; \; 
        \sigma^2=\frac{\pi^2}{6a^2},
\label{Eq:eq3}
\end{equation}
where $0.577 \approx -\Gamma'(1)$ is Euler's constant, $0.367 
\approx ln(ln(2))$ and $\sigma$ is the standard deviation of the 
extremes. The standard form for the Gumbel, $F(x)$ and $f(x)$, 
is shown in Fig. 1. Note that for BCGs, we will be considering
minimum extremes (because more negative magnitudes are brighter) 
and the curves will be inverted ($x \to -x$). Henceforth, we will call 
$f(x)$ the Gumbel distribution.

\begin{figure}
\vbox
{\epsfig{file=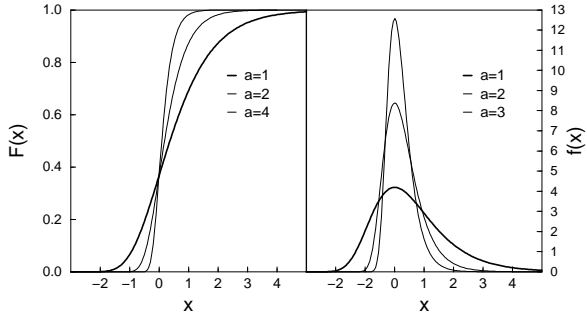,height=4.5cm}}
\caption{The Gumbel distribution for maximum extremes.}
\label{figure1}
\end{figure}

\section{Brightest Cluster Galaxies}
\subsection{Past Results}
Researchers have described BCGs as special, statistical extremes of a 
normal population and a mixture of the two (Peebles 1968; Peach 1969; 
Sandage 1972, 1976; Bhavsar \& Barrow 1985; Bhavsar 1989; 
Postman \& Lauer 1995).

The motivation for proposing that BCGs are special is due to the 
small dispersion observed in BCG magnitudes (Peach 1969; 
Sandage 1972, 1976). These authors argue that such a small 
dispersion is not sufficiently explained by the steepness of the 
luminosity function. In addition, astronomers observe a class of 
BCGs that are morphologically different, called $cD$ galaxies. 
These galaxies are giant ellipticals and often have features, such 
as multiple nuclei and large envelopes, that distinguish them 
from normal galaxies.

On the other hand, Peebles (1968) argues that BCGs are just the 
extreme tail-end of normal galaxies that form in clusters via some 
stochastic process. In this case, the brightest galaxy in a given 
cluster is simply the brightest normal galaxy and, therefore, the 
distribution of BCG magnitudes is a Gumbel. (It is interesting to 
note that Peebles, independently of Fisher \& Tippett, derived 
the Gumbel distribution for BCGs for the special case of an 
exponential luminosity function.)

Bhavsar (1989) contends neither of these scenarios adequately 
describes the observed distribution of BCG magnitudes and 
argues for a mixed population. Suppose that a special class of 
Galaxies exists but that not all clusters have a special galaxy. 
In clusters with no special galaxy, the BCG is simply the 
brightest normal galaxy. In a cluster containing at least one 
special galaxy, either all the normal galaxies are fainter, or 
the brightest normal galaxy(ies) out-shines the special one(s) 
and attains the first rank. For these reasons, one might expect 
both types of galaxies to comprise the BCG population. In 
what follows, we investigate these assumptions quantitatively 
by analyzing Lauer \& Postman's (1994) data set and revisiting 
the one used by Bhavsar (1989).

\subsection{The Distribution of BCG magnitudes}
In the case of one population, the distribution function is 
straight forward. If BCGs are all drawn from a special class 
of objects, it has been assumed that BCG magnitudes are 
normally distributed (Peach 1969; Sandage 1972, 1976;
Postman \& Lauer 1995). In this case, referred to henceforth 
as model A, the probability distribution of special galaxies, 
$f_{sp}$, is a Gaussian, $f_{g}$, with mean $M_{g}$, 
standard deviation $\sigma$ and normalization such that the 
integral over all magnitudes, $M$, is unity. The distribution
function is as follows:
\begin{equation}
        f_{sp}(M)=f_{g}=\frac{1}{\sigma \sqrt{2\pi}}
        e^{-\frac{(M-M_{g})^2}{2 \sigma^2}}.
\label{Eq:eq4}
\end{equation}
If BCGs are simply the brightest of a normal set of galaxies
(Peebles 1968), henceforth referred to as model B, the 
probability distribution of their magnitudes, $f_{nor}$, is a 
Gumbel, $f_{G}$, given by Equation (2), with $x \to -M$ and 
$x_0 \to M^* = M_{G}+\frac{0.577}{a}$ (Bhavsar \&
Barrow 1985):
\begin{equation}
        f_{nor}(M)=f_{G}=ae^{a(M-M^*)-e^{a(M-M^*)}},
\label{Eq:eq5}
\end{equation}
where $M_{G}$ is the mean of the extremes and $a$ is a measure 
of the steepness of fall of the parent distribution.

In the case of two populations (Bhavsar 1989), we derive the 
distribution that $M$ should have from the contributions of the 
two individual populations. Consider $N$ clusters of galaxies 
and suppose that $n<N$ have at least one special galaxy. Let the 
independent magnitude-distribution of normal and special galaxies, 
respectively, be $f_{nor}$ and $f_{sp}$. The total 
magnitude-distribution function, $f_{tot}$, is then given by:
\begin{eqnarray}
        f_{tot}(M)&=&d \cdot [f_{sp}\cdot\int_M^\infty f_{nor}(M') dM'+ \nonumber\\
                    &&f_{nor}\cdot\int_M^\infty f_{sp}(M') dM']+ \nonumber\\
                    &&(1-d)f_{nor},
\label{Eq:eq6}
\end{eqnarray}
where $d = n/N$. The first (second) term is the probability of picking 
a special (normal) galaxy, with absolute magnitude $M$, from a 
cluster containing both populations with the condition that all the 
normal (special) galaxies are fainter. The third term gives the 
probability of picking a galaxy, with absolute magnitude $M$, in 
clusters containing only normal galaxies. Equation (6) is true for 
all well-behaved functions $f_{nor}$ and $f_{sp}$. If $f_{nor}$ and 
$f_{sp}$ are normalized to unity, then so is the resulting total 
distribution function $f_{tot}$. (Note that Equation (6) works, in 
general, whenever there are two independent populations competing 
for first rank.)

For BCGs, we consider three different two-population models. The 
first is the case discussed above with the brightest normal galaxies 
comprising one population and a special class of galaxies comprising 
the other. We call this case model C and write the total distribution as 
$f_{Gg}$ (where `$Gg$' stands for `{\bf$G$}umbel + {\bf$g$}aussian'). 
To obtain the final form of $f_{Gg}$ we note that:
\begin{equation}
        I_G=\int_M^\infty f_{G}(M') dM'=F(M),
\label{Eq:eq7}
\end{equation}
where $F(M)$ is given by Equation (1) with $x \to -M$, and
$x_0 \to M_{G}+\frac{0.577}{a}$.
Second, we note that:
\begin{equation}
        I_g=\int_M^\infty f_{g}(M') dM'=(1 \pm erf|M-M_{g}|)/2,
\label{Eq:eq8}
\end{equation}
where $erf$ is the error function. The upper sign is for $M < M_{g}$ 
and the lower sign is for $M > M_{g}$. Thus, we may rewrite 
$f_{Gg}$ by substituting in $I_G$ and $I_g$:
\begin{equation}
        f_{Gg}(M)=d \cdot [f_{g} \cdot I_G+f_{G} \cdot I_g]+(1-d)f_{G}.
\label{Eq:eq9}
\end{equation}
Other possible combinations of assigning $f_G$ and $f_g$ to the two 
populations result in models D and E. In the case of model D, both 
distributions are Gaussian and the total distribution function, $f_{gg}$, 
is given by:
\begin{equation}
        f_{gg}(M)=d \cdot [f_{g2} \cdot I_{g1}+f_{g1} \cdot I_{g2}]+(1-d)f_{g1},
\label{Eq:eq10}
\end{equation}
where the notation is self-evident and the two Gaussians are 
characterized, respectively, by $M_{g1}$, $\sigma_1$ and 
$M_{g2}$, $\sigma_2$. In the case of model E, both 
distributions are Gumbels ($f_{sp}$ is also a Gumbel) and 
the total distribution function, $f_{GG}$, is given by:
\begin{equation}
        f_{GG}(M)=d \cdot [f_{G2} \cdot I_{G1}+f_{G1} \cdot I_{G2}]+(1-d)f_{G1},
\label{Eq:eq11}
\end{equation}
where the two Gumbels are characterized, respectively, by $M_{G1}$, 
$a_1$ and $M_{G2}$, $a_2$. Table 1 summarizes the forms of the 
five models.

\begin{table}
\caption{Distribution components for the five models.}
\vspace{2mm}
\centerline{\begin{tabular}{ccc} 
 MODEL & $f_{nor}$ & $f_{sp}$ \\ 
\\
 A & -- & $f_{g}$ \\ 
 B & $f_{G}$ & -- \\ 
 C & $f_{G}$ & $f_{g}$ \\ 
 D & $f_{g1}$ & $f_{g2}$ \\
 E & $f_{G1}$ & $f_{G2}$ \\ 
\end{tabular}}
\end{table}	

\section{Modeling the Data}
\subsection{Data Sets}
We utilize two data sets from the literature. First, we reanalyze the 
data used by Bhavsar (1989). This is a 93 member subset of 116 
metric BCG visual-intrinsic (VI) magnitudes compiled by Hoessel, 
Gunn \& Thuan (1980), henceforth referred to as ``HGT''. These 93 
are the data from clusters of richness 0 and 1 only; Bhavsar ignores 
the rest of the BCGs in order to keep the data set homogeneous. 
The BCG magnitudes are internally consistent to 0.04 magnitudes,
as published in HGT. Second, we analyze the 119 metric BCG 
magnitudes, taken in the Kron-Cousins $R_c$ band, compiled by Lauer 
\& Postman (1994), henceforth referred to as ``LP''. The data were 
corrected for local and possible large scale galactic motions. The 
119 LP data are comprised of BGCs from 107 clusters of  richness 
0 \& 1, and 9, 2 and 1 of richness 2, 3 and 4, respectively (Abell,
Corwin \& Olowin 1989). We find that removing the 12 BCGs 
from clusters of richness class $\geq$ 2 does not significantly change 
the distribution of the LP data. This is consistent with Sandage's (1976) 
result that BCG magnitude is independent of cluster-richness. The 
internal consistency of the set is 0.014 magnitudes, as published in 
Postman \& Lauer (1995). Bhavsar (1989) proposes a two-population 
model for the HGT data. His maximum-likelihood fit is consistent with 
the data and has parameter-values consistent with physically measured 
quantities. Postman \& Lauer (1995) conclude that the LP data are
consistent with a Gaussian. 

There are differences in the data sets that could be the reason for the 
disagreement between Bhavsar (1989) and Postman \& Lauer (1995). 
The two were obtained in different optical bands. The mean of the 
HGT data set is 0.2 magnitudes brighter than the mean of the LP data 
set. The two data sets have 34 galaxies in common. Comparing the 
subset of 34, we find that the HGT values are, on average, 0.06 $\pm$ 
0.19 magnitudes brighter than the LP values. A two-sample 
Kolmogorov-Smirnov (K-S) Test addresses the consistency of the two data sets in 
describing the same population of objects. The null hypothesis is that 
the same distribution describes both data sets. We find that the two data 
sets fail the null hypothesis at the 82\% confidence level. Therefore, we 
do not expect the same parameters or distribution to describe both sets. 
These discrepancies may need further investigation, but such an analysis 
is outside the scope of this work. We investigate each data set separately 
and present our results.

\subsection{Fitting Method}
We consider models A-E discussed above. The two-population 
distributions have five parameters each: two means, two standard 
deviations and the fraction, $d$, of clusters that contain a special 
population of galaxies. If there is no population of special galaxies, 
then $d$ = 0. We use maximum-likelihood fitting.  The theory behind 
this method is discussed in Press, et al. (1992). The Maximum-Likelihood 
fit to a data set of size $N$ for a function, $f$, are the parameters, 
$\bf a$, that maximize the likelihood function:
\begin{equation}
          L=\prod_{i=1}^N f(x_i;\bf a),
\label{Eq:eq12}
\end{equation}
where the $f(x_i;\bf a)$ are the values of the probability density, $f$, 
evaluated at each of the $N$ data points, $x_i$. For a certain $f$,
one finds the set of parameters that maximizes the product, $L$.

\section{Results}
\subsection{Parameters and Fits}
After obtaining parameters from the maximum-likelihood method for 
models A-E for both data sets, we compute the K-S statistics. We list 
the results in Tables 2 \& 3, respectively. Lower values of the K-S 
$D$-statistic correspond to lower values of rejection probability, $P$, 
and thus denote a better fit. Figs. 2 \& 3 illustrate the performance of 
each of the five models. Note that the distributions use the parameters 
obtained by the maximum-likelihood method, using $every$ data point, 
and are $not$ a fit to the particular histograms.

\begin{table}
\caption{Fit-parameters for the HGT data for models A-E.}
\centerline{\begin{tabular}{cc} 
 MODEL A & MODEL B \\ 
 $M_{g}$=-22.63 & $M_{G}$=-22.66 \\   
 $\sigma$=0.34 & $a$=2.82 \\   
 $D$=0.0876 & $D$=0.1174 \\  
 $P$=0.531 & $P$=0.848 \\ 
\end{tabular}}
\vspace{4mm}
\centerline{\begin{tabular}{ccc} 
 MODEL C & MODEL D & MODEL E \\ 
 $M_{G}$=-22.30 & $M_{g1}$=-22.29 & $M_{G1}$=-22.40 \\   
 $M_{g}$=-22.79 & $M_{g2}$=-22.83 & $M_{G2}$=-22.86 \\   
 $d$=0.64 & $d$=0.62 &  $d$=0.48 \\   
 $a$=4.11 & $\sigma_1$=0.24 & $a_1$=3.70 \\   
 $\sigma$=0.20 & $\sigma_2$=0.19 & $a_2$=8.83 \\   
 $D$=0.0562 & $D$=0.0519 & $D$=0.0525 \\  
 $P$=0.063 & $P$=0.032 & $P$=0.036 \\ 
\end{tabular}}
\end{table}

\begin{table}
\caption{Fit-parameters for the LP data for models A-E.}
\centerline{\begin{tabular}{cc} 
 MODEL A & MODEL B \\ 
 $M_{g}$=-22.43 & $M_{G}$=-22.45 \\ 
 $\sigma$=0.33 & $a$=2.99 \\ 
 $D$=0.0565 & $D$=0.1173 \\  
 $P$=0.162 & $P$=0.926 \\  
\end{tabular}}
\vspace{4mm}
\centerline{\begin{tabular}{ccc} 
 MODEL C & MODEL D & MODEL E \\ 
 $M_{G}$=-21.84 & $M_{g1}$=-22.11 & $M_{G1}$=-22.18 \\    
 $M_{g}$=-22.44 & $M_{g2}$=-22.52 & $M_{G2}$=-22.52 \\       
 $d$=0.95 & $d$=0.72 & $d$=0.64 \\    
 $a$=5.11 & $\sigma_1$=0.26 & $a_1$=3.65 \\   
 $\sigma$=0.32 & $\sigma_2$=0.30 & $a_2$=5.33 \\  
 $D$=0.0570 & $D$=0.0527 & $D$=0.0421 \\  
 $P$=0.158 & $P$=0.098 & $P$=0.014 \\  
\end{tabular}}
\end{table}

\begin{figure}
\vbox
{\epsfig{file=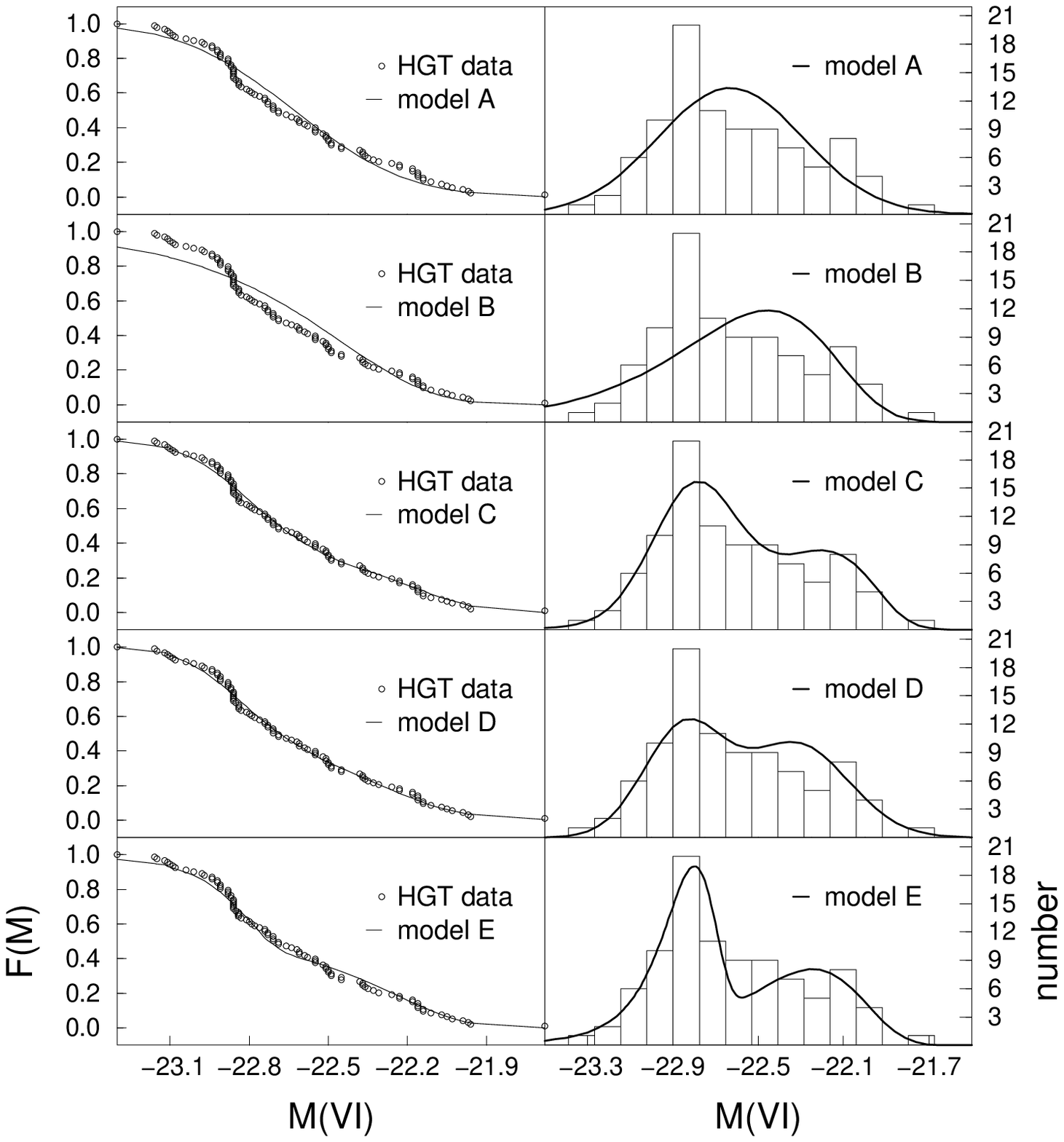,height=9.0cm}}
\caption{Left-hand column shows cumulative distribution function for the 
 HGT data and the maximum-likelihood fits for each of the five models. 
 Right-hand column shows HGT histogram with a plot of the differential
 distribution for each
 of the five models.}
\label{figure2}
\end{figure}

\begin{figure}
\vbox
{\epsfig{file=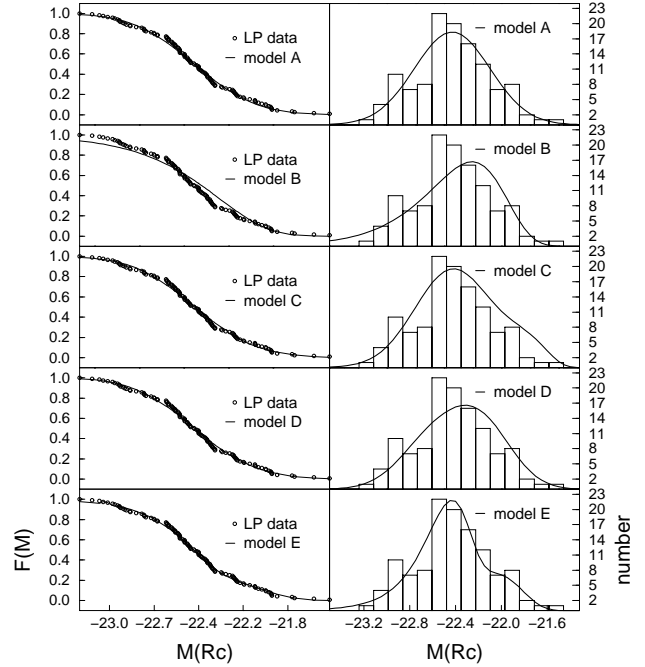,height=9.0cm}}
\caption{Left-hand column shows cumulative distribution function for the 
 LP data and the maximum-likelihood fits for each of the five models. 
 Right-hand column shows LP histogram with a plot of the differential 
 distribution for each of the five models.}
\label{figure3}
\end{figure}

\subsection{Comparison with Previous Work}
We compare our results with Bhavsar (1989) and Postman \& Lauer (1995). 
Bhavsar's (1989) two-population model is our model C. He 
uses maximum-likelihood fitting and his best-fitting parameters are 
$M_{G}$ = -22.31, $M_{g}$ = -22.79, $d$ = 0.63, $a$ = 4.01 and 
$\sigma$ = 0.21. Our parameters are in excellent agreement. Minor 
variation is expected due to differences in fitting techniques. Postman 
\& Lauer (1995) argue against Bhavsar's two-population model and 
claim that BCG magnitudes are Gaussian, based on a 26\% confidence 
level.

In agreement with both Bhavsar (1989) and Postman \& Lauer (1995), 
it is clear from Tables 2 \& 3 and Figs. 2 \& 3 that for both data sets no 
Gumbel distribution describes the BCG data. This rejects the Gumbel 
hypothesis (model B) with 85\% and 93\% confidence levels, respectively, 
for the HGT and LP sets. For the HGT data, the Gaussian fails at the 53\% 
confidence level, while for the LP data, the rejection confidence is 16\%. 
The difference between our value of 16\% and Postman \& Lauer's value 
arises because our result is for the maximum-likelihood fit Gaussian, 
while Postman \& Lauer's is for a Gaussian with the same mean and 
standard deviation as the LP data.

The relatively high rejection-confidence of the one-population models 
has motivated us to investigate two-population models. The presence of 
cD galaxies strongly suggests the possibility of another population. 
Overall, the two-population models fit the data much better than do the 
Gumbels and as well or better than do the respective Gaussians. The 
larger number of parameters is taken into account by the statistical 
estimators when calculating the confidence of  rejecting the null 
hypothesis. Moreover, the parameters are physical quantities that 
are observationally verifiable (Bhavsar 1989).

Our result that no one model or set of parameters describes both 
data sets is consistent with the fact that a two-sample K-S Test 
indicates that the sets are not consistent with one another. Postman 
\& Lauer (1995) have raised questions regarding HGT's BCG 
classification and sky subtraction.

\subsection{Physical Motivation}
Researchers have suggested various mechanisms whereby a second 
population with a brighter average metric magnitude could evolve 
from the bright normal galaxies. Cannibalism, the process by which 
large galaxies in the central regions of rich clusters grow at the 
expense of smaller galaxies (Ostriker \& Hausman 1977; Hausman 
\& Ostriker 1978), is one possibility. The existence of giant elliptical 
and cD galaxies near the centre of approximately half of all rich 
clusters supports this hypothesis. These galaxies always lie at the
tail-end of their cluster-luminosity functions. The occurrence of 
cannibalism continues to be debated (Merritt 1984).

Motivated by the existence in the literature of strong arguments for 
such a process, we build a very simple schematic to study its 
$statistical$ effects on the population of first-ranked galaxies. We 
make two assumptions: (i) at an early epoch the BCGs all belonged 
to one population and (ii) galaxies from the bright end of this 
population evolve, resulting in a random boost to their luminosity. 
We construct a set of $N$ galaxies with an exponential luminosity 
function between absolute magnitudes -22.0 and -23.0. This 
represents the galaxies at the bright end of cluster luminosity 
functions that are candidates for a boost. A random number, $n$, 
of these galaxies undergo a random boost between 0.1 and 0.9 
magnitudes. We label the boosted subset as $n_b$. We choose 
this range for the following reasons. First, Hausman \& Ostriker 
(1978) show via a simulation that one would expect a large galaxy 
to gain, on average, 0.5 magnitudes during its first cannibalistic 
encounter. This is consistent with Aragon-Salamanca, Baugh \& 
Kauffmann (1998), who state that BCGs were approximately 0.5 
magnitudes fainter at $z$ = 1.  Second, we limit ourselves to one 
encounter because Merritt (1984) argues that the time scale for 
galactic encounters is too long for cannibalism to be common in 
the universe. We wish to investigate the magnitude-distribution 
of the resulting boosted population. These represent the special 
galaxies mentioned previously. Specifically, this distribution 
could give us insight into the form of $f_{sp}$.

To our surprise, we find that the distribution, $f_{sp}$, of $n_b$ 
is a Gumbel! The K-S Test rejects the Gaussian hypothesis at the 
98\% confidence level. Conversely, the Gumbel distribution, with 
the same mean and deviation as the data, fits well, with only a 7\% 
confidence level for rejection. We summarize these results in Table 
and Fig. 4. Thus, the two-population model E (a combination of two 
Gumbels), which is best-fitting for the newer LP data, has a 
physical basis. 

\begin{table}
\caption{Fit-Parameters for the $n_b$ data.}
\vspace{2mm}
\centerline{\begin{tabular}{cc} 
 GAUSSIAN FIT & GUMBEL FIT  \\ 
 $M_{g}$=-22.64 & $M_{G}$=-22.64 \\ 
 $\sigma$=0.28 & $a$=4.58 \\ 
 $D$=0.0827 & $D$=0.0296 \\ 
 $P$=0.978 & $P$=0.067 \\ 
\end{tabular}}
\end{table}

\begin{figure}
\vbox
{\epsfig{file=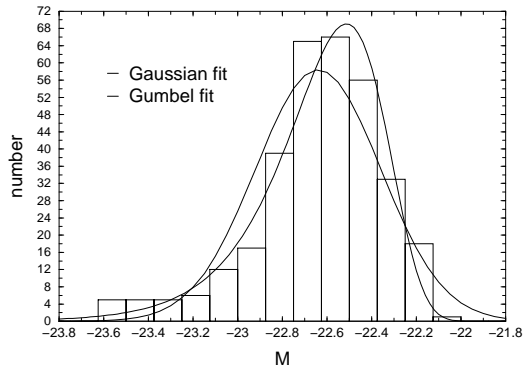,height=5.0cm}}
\caption{The end-result of boosting an initial exponential luminosity
 function compared with a Gaussian and a Gumbel.}
\label{figure4}
\end{figure}

\section{Conclusion}
For more than thirty years, cosmologists have debated the nature
of the magnitude-distribution of brightest cluster galaxies. Peebles (1968) 
and Sandage (1972, 1976) \& Peach (1969) reach markedly different 
conclusions. More recently, Bhavsar (1989) and Postman \& Lauer 
(1995) differ regarding the population(s) that comprise the first-ranked 
galaxies. In light of this controversy, we have conducted a new 
examination of the distribution of BCG magnitudes. We consider the 
BCGs as $a$ $class$ $of$ $objects$ to which we may apply well 
established results from extreme value theory. We find that there 
are a number of models that perform well in describing the HGT 
and LP data sets. Though a Gaussian fits both data sets, the 
confidence limits warrant further investigation of two-population 
models.

Tables 2 \& 3 clearly show that we should reject the Gumbel (model B) 
as a fit, i.e., the hypothesis that all BCGs are statistical extremes. The 
Gaussian (model A) is marginally acceptable but without physical basis. 
Two-population models, in particular, the  three combinations of $f_{G}$ 
and $f_{g}$, describe the data very  well. Tables 2 \& 3 show their relative 
merits. Model E stands out as giving the best overall fit and is motivated by 
a physical basis. Therefore, it is most likely that there are two populations 
of BCGs: the extremes of a normal population and a class of atypical 
galaxies with a brighter average mean.  \\\\

We thank Marc Postman for sending us the LP data. This research was 
supported by an ANN grant from the US Department of Education and 
the Kentucky Space Grant Consortium.

%\onecolumn
%\begin{table}
%\caption{Fit-parameters for the HGT data for models A-E.}
%\vspace{4mm}
%\begin{tabular}{ccccc} 
% MODEL A & MODEL B & MODEL C & MODEL D & MODEL E \\ 
%\\
% $M_{g}$=-22.63 & $M_{G}$=-22.66 & $M_{G}$=-22.30 & $M_{g1}$=-22.29 & $M_{G1}$=-22.40 \\   
% -- & -- & $M_{g}$=-22.79 & $M_{g2}$=-22.83 & $M_{G2}$=-22.86 \\   
% -- & -- & $d$=0.64 & $d$=0.62 &  $d$=0.48 \\   
% $\sigma$=0.34 & $a$=2.82 & $a$=4.11 & $\sigma_1$=0.24 & $a_1$=3.70 \\   
% -- & -- & $\sigma$=0.20 & $\sigma_2$=0.19 & $a_2$=8.83 \\   
% $D$=0.0876 & $D$=0.1174 & $D$=0.0562 & $D$=0.0519 & $D$=0.0525 \\  
% $P$=0.531 & $P$=0.848 & $P$=0.063 & $P$=0.032 & $P$=0.036 \\ 
%\end{tabular}
%\end{table}

%\begin{table}
%\caption{Fit-parameters for the LP data for models A-E.}
%\vspace{4mm}
%\begin{tabular}{ccccc} 
% MODEL A & MODEL B & MODEL C & MODEL D & MODEL E \\ 
%\\
% $M_{g}$=-22.43 & $M_{G}$=-22.45 & $M_{G}$=-21.84 & $M_{g1}$=-22.11 & $M_{G1}$=-22.18 \\    
% -- & -- & $M_{g}$=-22.44 & $M_{g2}$=-22.52 & $M_{G2}$=-22.52 \\       
% -- & -- & $d$=0.95 & $d$=0.72 & $d$=0.64 \\    
% $\sigma$=0.33 & $a$=2.99 & $a$=5.11 & $\sigma_1$=0.26 & $a_1$=3.65 \\   
% -- & -- & $\sigma$=0.32 & $\sigma_2$=0.30 & $a_2$=5.33 \\  
% $D$=0.0565 & $D$=0.1173 & $D$=0.0570 & $D$=0.0527 & $D$=0.0421 \\  
% $P$=0.162 & $P$=0.926 & $P$=0.158 & $P$=0.098 & $P$=0.014 \\  
%\end{tabular}
%\end{table}

\label{lastpage}

\end{document}